\documentclass[12pt,preprint]{aastex}

\usepackage{psfig}
\newcommand{\p}{\hbox{$^\prime$}}

\shorttitle{VLT adaptive optics observations of Circinus}
\shortauthors{Prieto et al.}

\begin{document}

\title{Unveiling the central parsec region of an AGN: the Circinus
  nucleus in the near infrared with the VLT}
\author{M. Almudena Prieto (1), K. Meisenheimer (1) }
\email{prieto@mpia.de}
\author{Olivier Marco (2) }
\author{Juha Reunanen (3), Marcella Contini (4), Y. Clenet (5), R.I. Davies (6), D. Gratadour (5), Th. Henning (1),  U. Klaas (1), J. Kotilainen (3),
Ch. Leinert (1), D. Lutz (6), D. Rouan (5),  N. Thatte (7) }

\affil{ (1) Max-Planck-Institut f\"ur Astronomie, Heidelberg,  Germany}
\affil{ (2) ESO Paranal, Santiago, Chile}
\affil{ (3) Tuorla Observatory, Finland}
\affil{ (4) Tel Aviv University, Israel}
\affil{ (5) Observatoire de Paris, France}
\affil{ (6) Max-Planck-Institut f\"ur extraterrestrische Physik,
  Garching,  Germany}
\affil{ (7) University of Oxford, UK}

\begin{abstract}

VLT J- to M\p-band adaptive optics observations of the Circinus Galaxy
on parsec scales resolve a central bright Ks-band source with a FWHM
size of 1.9 $\pm$ 0.6 pc.  This source is only visible at wavelengths
longward of 1.6\,$\mu$m and coincides in position with the peak of the
[Si VII]~2.48\,$\mu$m coronal line emission.  With respect to the peak
of the central optical emission, the source is shifted by $\sim$
0.15\arcsec\ (2.8 pc) to the south-east.  Indeed, it defines the
vertex of a fairly collimated beam which extends for $\sim$ 10 pc, and
which is seen in both continuum light shortward of 1.6\,$\mu$m and in
H$\alpha$ line emission.  The source also lies at the center of a
$\sim$ 19 pc size [Si VII] ionization {\it bicone}.

Identifying this source as the nucleus of Circinus, its size is
compatible with a putative parsec-scale torus.  Its spectral energy
distribution, characterized by a prominent narrow peak, is compatible
with a dust temperature of 300 K.  Hotter dust within a 1 pc radius of
the center is not detected.  The AGN luminosity required to heat this
dust is in the range of X-ray luminosities that have been measured
toward the central source.  This in turn supports the existence of
highly obscuring material, with column densities of
$10^{24}$\,cm$^{-2}$, that must be located within 1 pc of the core.



\end{abstract}


\section{Introduction}

Circinus is the second closest AGN ($\sim$4~Mpc, Freeman et al. 1977;
$1\arcsec\sim19$~pc ) after Centaurus~A in the Southern Hemisphere.
It is a SA(s)b galaxy inclined by $\sim \rm65 \deg$ with a Seyfert
type 2 AGN at its center. 
Several dust lanes running north to south across the galaxy hide a large
fraction of its eastern side. 
Probably for this reason, Circinus shows a one-sided cone of ionized
gas which extends to the north-west in H$\alpha$ and [OIII] images on
kpc scales (Marconi et al. 1994), and in X-rays on smaller scales
(Smith \& Wilson 2001).
The hard 1-100\,keV X-ray spectrum indicates the presence of an obscured
nucleus with N(H) $\sim 4\times10^{24}$\,cm$^{-2}$ (Matt et al. 1999).
The highest spatial resolution view of its central region comes from
VLBI maps of H$_2$O maser emission (Greenhill et al 2003). 
The H$_2$O emission is interpreted  in terms of a central
edge-on Keplerian disk with inner and outer radii of 0.1 and
0.4~pc respectively, and outflowing material up to 1~pc from the
center.



This paper presents near-to-diffraction limited images of the Circinus
galaxy in the 1.2--5\,$\mu$m range observed with the adaptive optics
assisted imaging-spectrograph NACO at the {\em Very Large Telescope}
(VLT).  
Circinus was observed as part of the multi-resolution imaging program
ZIDNAG (Zooming Into Dusty Nuclei of Active Galaxies)
aimed at imaging the dusty tori in the closest AGN with both NACO
and, subsequently, VLTI interferometric observations at
10\,$\mu$m and 2.2\,$\mu$m.


The physical scales associated with the spatial resolution of the
Circinus images presented in this paper 
lie in the range 1.5--3 pc, and hence probe the presence and
dimensions of obscuring material -- the putative torus -- around this
type 2 AGN.

\section{Observations and analysis}

NACO images in broad J (1.3\,$\mu$m), Ks (2.2\,$\mu$m), L\p\
(3.8\,$\mu$m), and  M\p\ (4.8\,$\mu$m) bands, and in narrow 2.42 and
2.48~$\mu$m bands, were
obtained with the UT4 unit telescope of the VLT in May 2003 (broad
band) and March 2003 (narrow band).
With the exception of the M\p-band image, which was taken in chopping
mode and has a reduced field of view of $12\arcsec\times12\arcsec$, 
all the images have a field of view of $27\arcsec\times27\arcsec$ and a
pixel scale of 0.027\arcsec\ per pixel.
The pixel scale fully samples the diffraction limit of the VLT in the Ks,
L\p, and M\p\ bands, but undersamples it at J.  
The wavelength of the narrow band image at 2.48\,$\mu$m is centered at
the position of the coronal [Si VII]~2.48\,$\mu$m line;
that at 2.42\,$\mu$m is an off-line image taken for continuum
subtraction.  

The optical wavefront sensor was used for all wavelengths.
Final images were made after sky-subtracting, shifting,
and co-adding a number of raw frames (about 30 in J and Ks; 170 and 90 in
L\p\ and M\p\ respectively) taken at random dithered positions within a
jitter box of $\sim20\arcsec$ about the central position of Circinus. 
The large number of raw frames and the size of the jitter width proved
to be a robust procedure for generating reliable sky images from the
science frames themselves without resorting to separate observations
of the sky.
Exposure times ranged from 3 min in J and Ks, and 5 min in L\p, to 
10 min in M\p, and 20 min in the narrow band filters.

The resulting FWHM spatial resolutions were measured
from several stars in the field of the science frames to be:
$0\farcs21\pm0\farcs01$ in the J-band,
$0\farcs16\pm0\farcs02$ in the Ks-band,
$0\farcs20\pm0\farcs01$ at 2.42\,$\mu$m, and
$0\farcs19\pm0\farcs01$ at 2.48\,$\mu$m. 
In the L\p\ and M\p\ bands, resolutions were derived from star images taken
after the Circinus observations, 
being $<0\farcs12$ and $<0\farcs13$ respectively.
The Strehl values achieved are difficult to asses as  Circinus nucleus
 is  extended, 
and  the stars in the sience frame that could be used for that purpose are are inmersed in the galaxy light. Nevertheless, for orientative purposes, the
 Strehl measured in real time 
by the Adaptive Optic system are provided: $\sim$ 20\% in J, 15\% in Ks and between  40- 50\% in the  L- and M-band images.

 
For comparison purposes, WFPC2
 broad-band 8140\,\AA\ (F814W, PC chip only) and 
 H$\alpha$ images, as well as NICMOS H-band (F160W)
 and 1.66\,$\mu$m narrow-band images were
 retrieved from the {\em Hubble Space Telescope} (HST) archive.

Fig. 1 presents a compilation of the most relevant images: 
from HST  F814W to NACO M\p-band, and [Si VII] 2.48\,$\mu$m ``pure'' line
emission image.  
The H, Ks, and 2.48\,$\mu$m images resemble
the 2.42\,$\mu$m image which already appears in Fig.~1 and so are not
shown.
The morphology of Circinus does not change at wavelengths shortward of
1.6\,$\mu$m.
However, at longer wavelengths its eastern side, which is largely
hidden by dust lanes in the F814W and J-band images, is progressively
revealed.
In the L\p- and M\p-bands, Circinus is reduced to a more compact
region (total diameter $\sim1\farcs2$) with very faint (S/N$\sim$3,
6\,mag fainter than the peak) extended emission in L\p. 
This is due to a combination of lower efficiency of the
NACO camera and the natural decrease in stellar flux at these
wavelengths.

\section{Astrometry of the Infrared Central Source}

Astrometric registration of the HST F814W  and NACO images was achieved
 thanks to the presence of several stars common to all images. For this 
purpose, the    F814W PC image, scale = 0.04'' per pixel  
was rebinned to that of  NACO. 
Based on the registration of  6--7 stars in the field, an
accuracy better than $\pm1$ pixel   was reached
 between the F814W, J, Ks, 2.42\,\micron, and 2.48\,\micron\ images;
 and about $\pm2$ pixels for the L\p-band image, for which only one
 star was available.
No registration was possible for the M\p-band image.

Following registration, it was found that the central peaks of the
F814W and J-band image coincided spatially.
Their position, however, is offset by $\sim0.15\arcsec$ (2.8 pc) north-west
 with respect to the central peak position found in all of the
 Ks, 2.42\,\micron, 2.48\,$\mu$m and  L\p images and also in the F160W
 image.
The morphology of the central emission is very different in the two cases.
The central emission in the F814W and J-band images appears diffuse
and rather collimated over a distance of $\sim0\farcs6$ (10\,pc) along
the inner main axis of Circinus' ionization cone (see the HST H$\alpha$
 image in Wilson et al. 2000). 
Close examination of the 1.66\,$\mu$m narrow-band ``continuum'' 
image shows a similar morphology although with less definition. 
On the other hand, the Ks, 2.42\,\micron\, 2.48\,$\mu$m and L\p,
 images show a bright compact source just 
located at the (south east) apex of the collimated beam seen in
 the F814W and J-band images (Fig. 2).  

The position of the infrared source also coincides with the peak emission of
 the (pure) [Si~VII]~2.48\,$\mu$m coronal line. 
Furthermore, the [Si VII]~2.48\,$\mu$m exhibits an elongated structure
 -- 19\,pc radius -- extended on both sides of the nucleus (Fig. 1).
The direction of the extended emission is not exactly along the main
 axis of the ionization cone, but it is certainly within
 the cone boundaries as defined at its base, as apparent from a
 comparison of Figs.~1 and~3.
Thus it appears that the [Si VII]~2.48\,$\mu$m emission, 
which is considerably less affected by extinction than H$\alpha$,
unveils the counter-cone of Circinus.
This result confirms the report of Maiolino et al. (2000) for
 counter-cone emission seen at rather low signal-to-noise in the 
[Si~VI]~1.96\,$\mu$m line.
Since [Si VII]~2.48\,$\mu$m probes Circinus' ionizing continuum
 at energies of about 200 eV, and considering the symmetric morphology
 of the emisison, it is reasonable to conclude that the central peak
 of [Si VII] emission pinpoints the active nucleus
 in Circinus.


Appraising all the evidence above, it appears that the infrared central
source detected at wavelengths longer than 1.6\,\micron\
qualifies as the true nucleus of Circinus.
At this location there is no enhanced emission in the F814W
or J-band images, 
indicating that this infrared central source is fully obscured at
 wavelengths $\lesssim1.3$\,$\mu$m.

An attempt to establish the absolute astrometry of the infrared central
 source was performed on the HST F814W image (PC chip)  with the USNO
 star catalog.
Based on the HST F814W pointing alone, the position of NACO Ks-band
 central source is found to be 
$14^h13^m9\fs96\pm0\fs02\ -$$65^\circ20\arcmin19\farcs91\pm0\farcs01$
 (with the uncertainties given here derived solely from the relative
 astrometry of the two images).
For the HST NICMOS F160W pointing, the position in right ascension is found
 the same but the declination is offset to the south by 0.44\arcsec.
For several stars identified in the HST F814W image (it was not
possible to locate any stars in the F160W image common to the USNO
catalog), 
the difference between the coordinates from the HST data and those
given in the USNO catalog were found to be smaller than the errors
reported for the USNO stars themselves, which were typically
0.9\arcsec\ on each axis.
We therefore assign an uncertainty of 0.9\arcsec\ to the absolute
position derived using the HST F814W (PC chip) astrometry.
We find thus that the position of Circinus IR source
 differs from the maser position
derived from VLBI (Greenhill et al. 2003), which is
$14^h13^m9\fs95\pm0\fs02\ -$$65^\circ20\arcmin21\farcs2\pm0\farcs1$
 (2000.0), by 1\farcs2, that is an additional 0\farcs3 beyond our
assigned error.
On the other hand, it is consistent with the 2MASS K-band position 
$14^h13^m9\fs92\ -$$65^\circ20\arcmin21\farcs57$ (quoted uncertainties
are approximately 0\farcs1).
Thus, a clear negative shift in declination applies to all
 the above optical/infrared positions with respect to the VLBI one.  
This offset may be related to an intrinsic
systematic shift between the radio and optical/infrared
coordinate reference systems.  
But if it were real, it could indicate that the maser source does not
actually pinpoint the nucleus but is instead part of the nuclear
H$_2$O outflow  measured by Greenhill et al.


\section{The Nature of the Infrared Central Source}

The morphology of the infrared central source in Circinus is rather
point-like at wavelengths longward of 1.6\,\micron; 
its FWHM, however, is larger than
that of the stars measured in the same images.
The size is best determined in the Ks-band and, after quadrature
subtraction of the resolution beam, is found to have a 
FWHM$=0\farcs11\pm0\farcs03$ ($1.9\pm0.6$ pc). 
In the L\p and M\p\ bands, the core sizes are (not deconvolved)
FWHM$=0\farcs185\pm0\farcs05$ and $0\farcs165\pm0\farcs05$ respectively.
 

In Table 1, the nuclear NACO fluxes in apertures of 0\farcs19
radius are provided, an aperture chosen to be about twice the
resolution achieved in the Ks-band. 
The core fluxes from the NICMOS H-band
and ESO/TIMMI2 N-band data in apertures less than 1\arcsec\ are
also included.  
Fig. 4 (black squares) shows  
the corresponding infrared spectral energy distribution
(SED) complemented by X-ray data from ROSAT (1 keV flux from Brinkmann et
al. 1994) and BeppoSAX (the 10 KeV flux is derived from the absorption
corrected nuclear luminosity,
F$_{\rm 2-10\,keV} = 3.4\times 10^{41}$\,erg\,s$^{-1}$ 
from Matt et al. 1999), and by radio data at various wavelengths and
in different apertures from the compilation of Contini et al. (1998).  
All the infrared measurements are associated with Circinus' central
infrared source except for the J-band, since the source is not
detected shortwards of this wavelength.
For consistency, the J-band value given is the integrated flux within
an aperture of radius 0\farcs19 measured at the position of the
infrared source.

The radio data, although of low
spatial resolution, is dominated by the Circinus nucleus (Elmouttie et
al. 1995);
the nuclear X-ray emission is resolved by Chandra into various
point-like sources surrounding a bright source which is associated with the
nucleus. 
These satellite sources have 2--10 keV unabsorbed
luminosities in the range $10^{37} - 10^{40}$\,erg\,s$^{-1}$ (Smith \&
Wilson 2001; no luminosity is provided for the nuclear source), from which
we infer that the more than one order of magnitud  flux by BeppoSAX  
originates mostly from the
nuclear source.


The most prominent feature in the SED is a narrow peak at infrared
wavelengths, indicative of dust emission in a narrow temperature
range.
This peak is 
well approximated by a simple modified black-body function with
temperature of about 300 K (Fig. 4, continuous line)
De-reddening for extinctions up to
A$_{\rm V}=20$ (see below) does not really modify the dust temperature,
mainly due to the steep shape of the infrared SED. 
Additional components of cooler dust may be present (this dust is not traced by the our data)
but hotter dust is not detected: it is either not present or
totally extincted.  
Some excess emission over the modified
black-body is apparent at 2.2\,$\mu$m and below but that may be due to
a combination of the underlying galaxy contribution and bremsstrahlung
emission from  cooling gas, both of which could still
contribute to these small apertures.
On the same Fig. 4, we  show a model of the complete 
radio to X-ray SED using the 
SUMA code (Contini et al. 2004 and references
therein). Briefly, this code considers the coupling effect of
photoionization -- from an external source, the AGN in this case -- 
and shocks into an ensemble of clouds randomly
distributed around the AGN and characterized by different physical
conditions of shock velocities, pre-shock
densities, preshock magnetic field, cloud geometrical thickness, 
and dust-to-gas ratios. As a preliminary result, we find the current  nuclear
 SED  to be characterized by shock-dominated clouds.
The infrared to  X-ray SED can be simultaneously 
fit by shock-dominated clouds with velocities $\gtrsim1000$\,km\,s$^{-1}$ 
and downstream densities n$_{\rm e} = 10,000$\,cm$^{-3}$. 
Both ranges of values are compatible 
with what one would expect at distances of few parsec radius from the
center. 
Accordingly, the infrared bump is dominated by dust reradiation from
clouds with those velocities while the X-ray to UV regime is dominated by
bremsstrahlung. 
The radio is pure synchrotron radiation generated by the Fermi mechanism
at the shock front. 
No radiation-dominated clouds appear in this model.  Since they must
exist, their absence is interpreted as an indication that heavy
obscuration is occurring at distances of less than 1 pc from the
nucleus.  We note that the contribution from these high velocity
clouds to the nebular spectrum is negligible, as found by Contini et
al. (1998).  These clouds are located very close to the center and at
most will contribute to the production of coronal and HeII
lines. Indeed, new high resolution spectra of optical and IR coronal
lines of Circinus reveal line widths with V$\sim 400 km~s{-1}$ 
(Rodriguez-Ardila, Prieto \& Viegas 2004) whereas
the average nebular line width measured in this galaxy is only 
100 $km~s{-1}$ . Independent
evidence for the existence of high velocity clouds is provided by the
maser outflow velocities of V$\sim 400 km~s{-1}$ measured at less than
0.1 pc from the center (Greenhill et al.)!  Larger maser velocities
may be present but a wider VLBI velocity baseline would be needed to
probe them.

Beyond the nuclear region of Circinus, the effect of dust obscuration is
best illustrated by the [J-2.42\,$\mu$m] color image (Fig.~\ref{jk}, 
a [J-Ks] color image shows exactly the same structure but the
2.42\,$\mu$m image is preferred as it includes continuum emission
only).  
Fig.~\ref{jk} shows a patchy distribution of bright and
dark regions, with the brighter regions located mostly to the south-east side
of the galaxy. 
These brighter regions correspond to stronger Ks-band emission, and 
spatially correlate with the darker zones seen, for example, in the
F814W and J-band images (Fig. 1).
Thus they most probably mark the location of cooler dust. 
Further comparison of the F814W and [J-2.42\,$\mu$m] images reveals
that this cool dust follows well defined channels that seem to spiral
down toward the center.

Comparison of the [J-2.42\,$\mu$m] colours at different points in the
galaxy to average values in normal ellipticals (J-K=0.95,
Giovanardi and Hunt, 1996) lead to extinction estimates in the
range A$_{\rm V}=2$--6\,mag (for a foreground screen) or 6--20\,mag
(if the dust and stars are mixed), depending on position. 


At the nucleus itself, the extinction is expected to be much higher
considering the Compton thick nature of the source.
Assuming a standard dust-to-gas ratio, a column density of
$10^{24}$\,cm$^{-2}$ translates into an extinction
of about 80 at 10\,$\mu$m, enough to dim any emission from the
central source even in the infrared.  
Since we do in fact see a 2-pc infrared central source, this
has to be subject to much less extinction. 
Considering its size and temperature, and assuming the
emission is due to dust reprocessing of radiation from the AGN, 
the inferred UV luminosity to heat this source
is $\sim 7\times10^{41}$\,erg\,s$^{-1}$. 
This is within the range of nuclear X-ray luminosities, 
L$_{\rm 2-10\,keV} \sim 3$--$17\times 10^{41}$\,erg\,s$^{-1}$, 
measured from the X-ray spectrum after correction for absorbing
columns of $10^{24}$\,cm$^{-2}$. 
Thus, the infrared central source in Circinus provides independent
support for the Compton thick nature of its core; it also indicates that
any  highly obscuring material should be located at distances less
than 1 pc from the core.
 
We note that the  SUMA modeling of the complete SED leads
to a different interpretation for the origin of the infrared bump. In 
this case,  the infrared bump is due to dust
heated primarily by shocks, whereas we found above that the
heating of the infrared central source is compatible with direct
illumination by the AGN.  While both possibilities are plausible, 
the current SED model is poorly constrained since it relies on insufficient 
information   from other spectral ranges, particularly the millimeter range. 
For an accurate modeling to be accomplished, further X-ray, far-infrared and
millimeter data with resolutions on scales of  a few parsecs are
needed.

\section{Conclusion}

The central source of Circinus at infrared wavelengths is resolved in
the Ks-band as a bright compact region with size $1.9\pm0.6$ pc (FWHM).
This source is only visible at wavelengths longward of 1.6\,$\mu$m, and it is
offset by 0.15\arcsec\ to the south-east of the brightest central 
emission seen in the J-band and HST optical images. 
The source is indeed located at the vertex of a rather collimated beam
($\sim10$ pc long) apparent at wavelengths shortward of 1.6\,$\mu$m. 
The direction and position of this collimated beam coincides
precisely with the main axis of Circinus' H$\alpha$ ionization cone.  
No collimation is seen at wavelengths beyond the K-band. 
As scattering is more effective at shorter
wavelengths, the observed feature may be nuclear light scattered by
the compact dusty structure in which the Circinus nucleus seems to
reside.

The source also coincides with the peak of the [Si VII]~2.48\,$\mu$m
 line emission; and with the center of a 19\,pc extended [Si VII]
 bicone of ionized gas.  On the basis of the arguments above, the
 infrared source that has been detected can be associated with the
 nucleus of Circinus.

The astrometric position of the infrared source differs by at least
0\farcs3 in declination from the VLBI maser position.  
While this may be due to a systematic difference between the
radio and optical/infrared reference frames, an alternative 
possibility is that the maser source is not related to the nucleus  
but instead to the outflowing material traced by the VLBI data.

The size of the infrared central source is compatible
 with a putative parsec-scale torus.
Its SED can be fully accounted for by dust at about 300 K, and the
 required AGN luminosity to heat such dust at a distance of 1 pc  
from the core is in the range of luminosities derived from X-ray data.
 This energy balance implicitly supports the presence of column
 densities as high as $10^{24}$\,cm$^{-2}$ at the nucleus.
The absorbing material has to be located at distances within 1 pc of
 the core and be rather inhomogeneously distributed in order that the
dust at 300\,K can have a direct view of the AGN.
Hotter dust is not detected at distance scales of 2 pc: it is either
not present or is heavily extincted. 
Additional cooler gas may certainly be present, probably distributed
 at slightly larger radial scales.
This cold dust is not sampled by our data, but
interferometric data at millimeter wavelengths should in the future 
 probe this regime.

Within a 200 pc radius of the central source in Circinus, the extinction
derived from color maps is inhomogeneous, with A$_{\rm V}=2$--6
depending on position. 
Close to the central source, higher extinctions of A$_{\rm V}=6$ (for
a foreground screen) or A$_{\rm V}=20$ (if the dust and stars are
mixed) are found.

\clearpage
\newpage

\begin{deluxetable}{lccc}
\tablecolumns{4}
\tablewidth{0pc}
\tablecaption{Photometry of Circinus. All data are from NACO except:  
N-band from ESO/TIMMI2 (Heijligers, 2003) and 
H-band from HST/NICMOS (Quillen et al. (2001)}
\label{sedTb}
\tablehead{\colhead{Band} &
\colhead{Magnitude} &
\colhead{Flux/mJy} &
\colhead{Aperture Diameter} }

\startdata
J            & 15     & $\lesssim$1.6 & 0\farcs38 \\
H            & 13.4   & 4.77          & 0\farcs1  \\
K            & 11.4   & 19 	      & 0\farcs38 \\
2.42\,$\mu$m & 10.56  & 31            & 0\farcs38 \\
L	     & 7.1    & 380 	      & 0\farcs38 \\
M	     & 4.8    & 1900          & 0\farcs38 \\
N            &        & 9700          & 1\arcsec  \\
$\rm [Si VII]$\,2.48\,\micron & & $2.0\times10^{-13}$\,erg\,cm$^{-2}$\,s$^{-1}$ & 2\farcs9 \\
\enddata

\end{deluxetable}
\clearpage
\newpage


\begin{figure}[t]



\centerline{\psfig{file=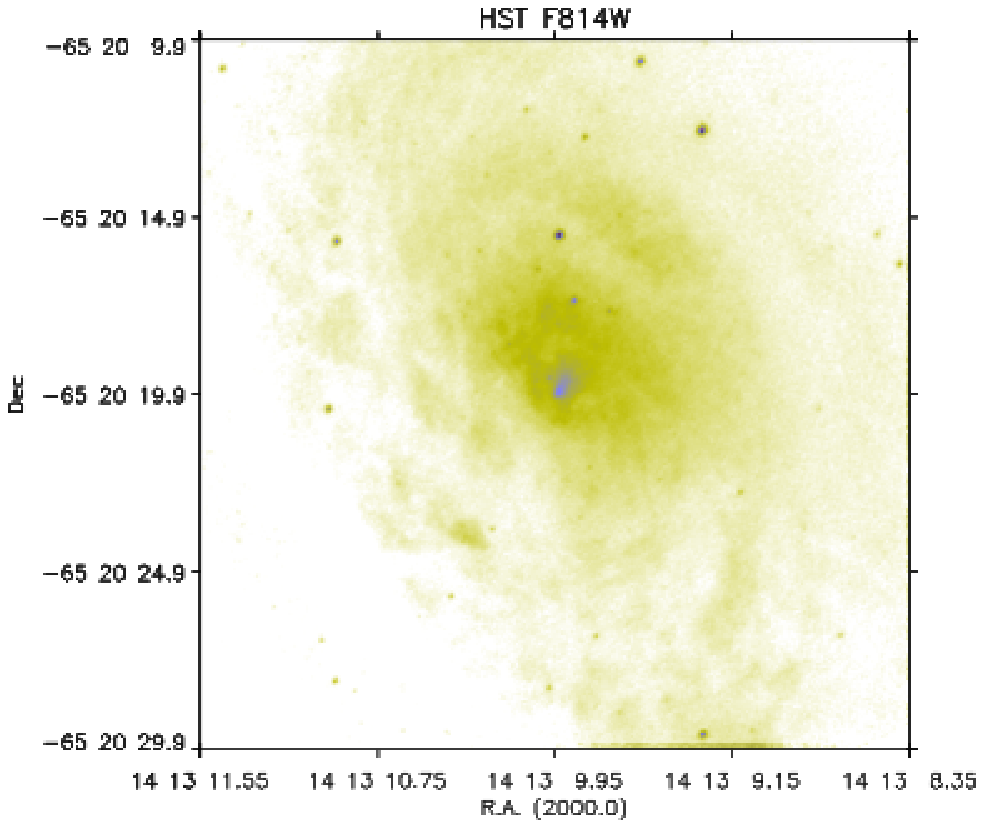,width=65mm}\hspace{5mm}\psfig{file=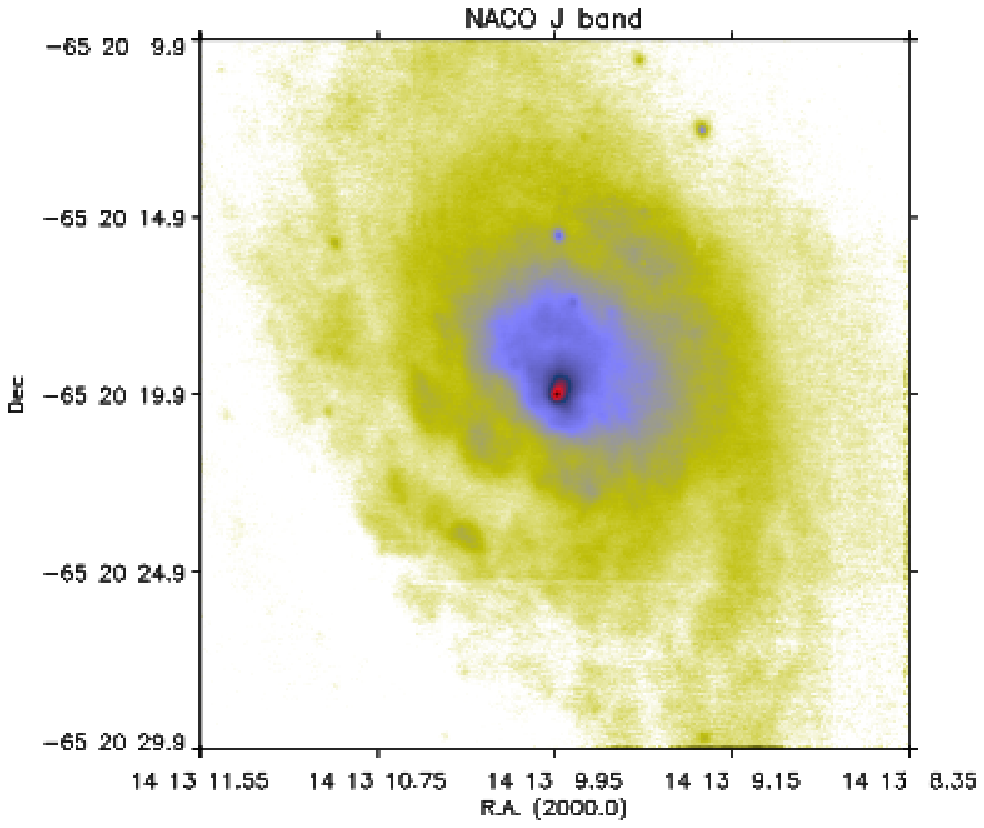,width=65mm}}
\centerline{\psfig{file=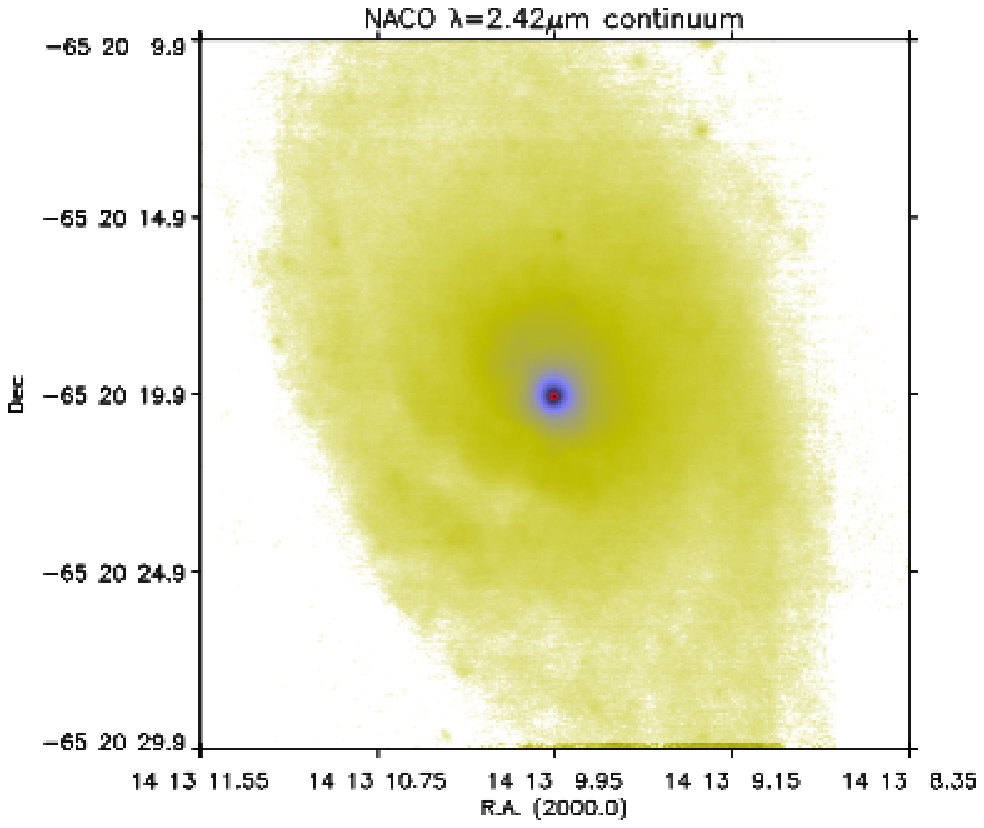,width=65mm}\hspace{5mm}\psfig{file=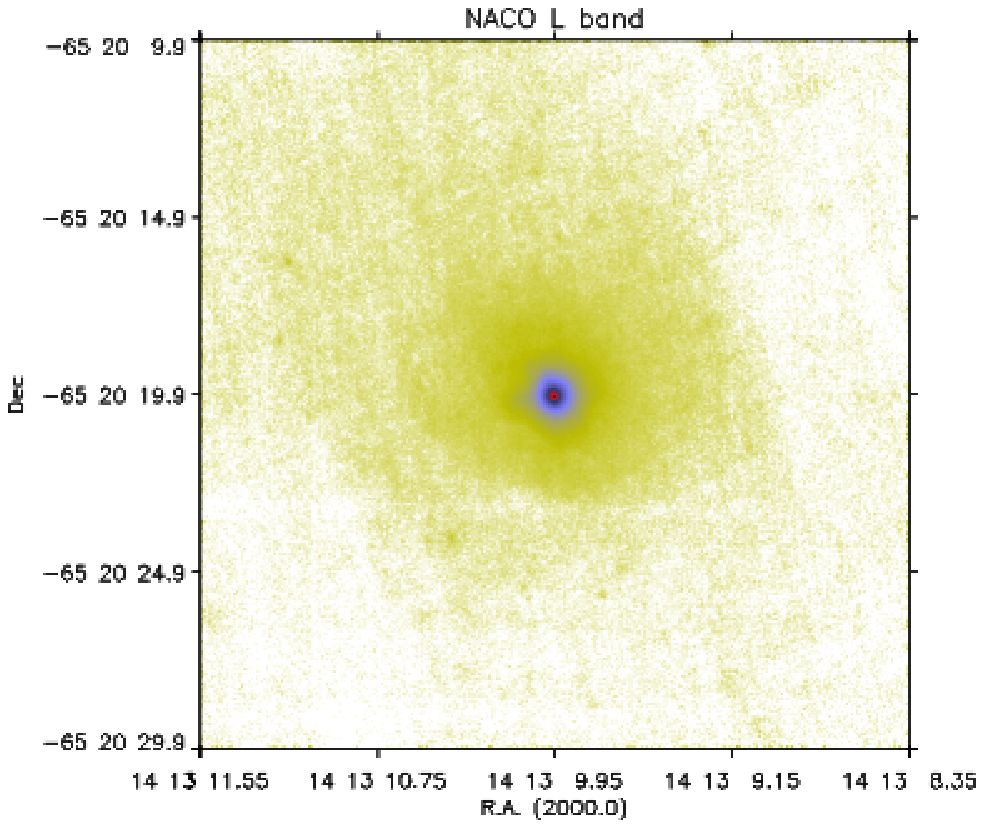,width=65mm}}
\centerline{\psfig{file=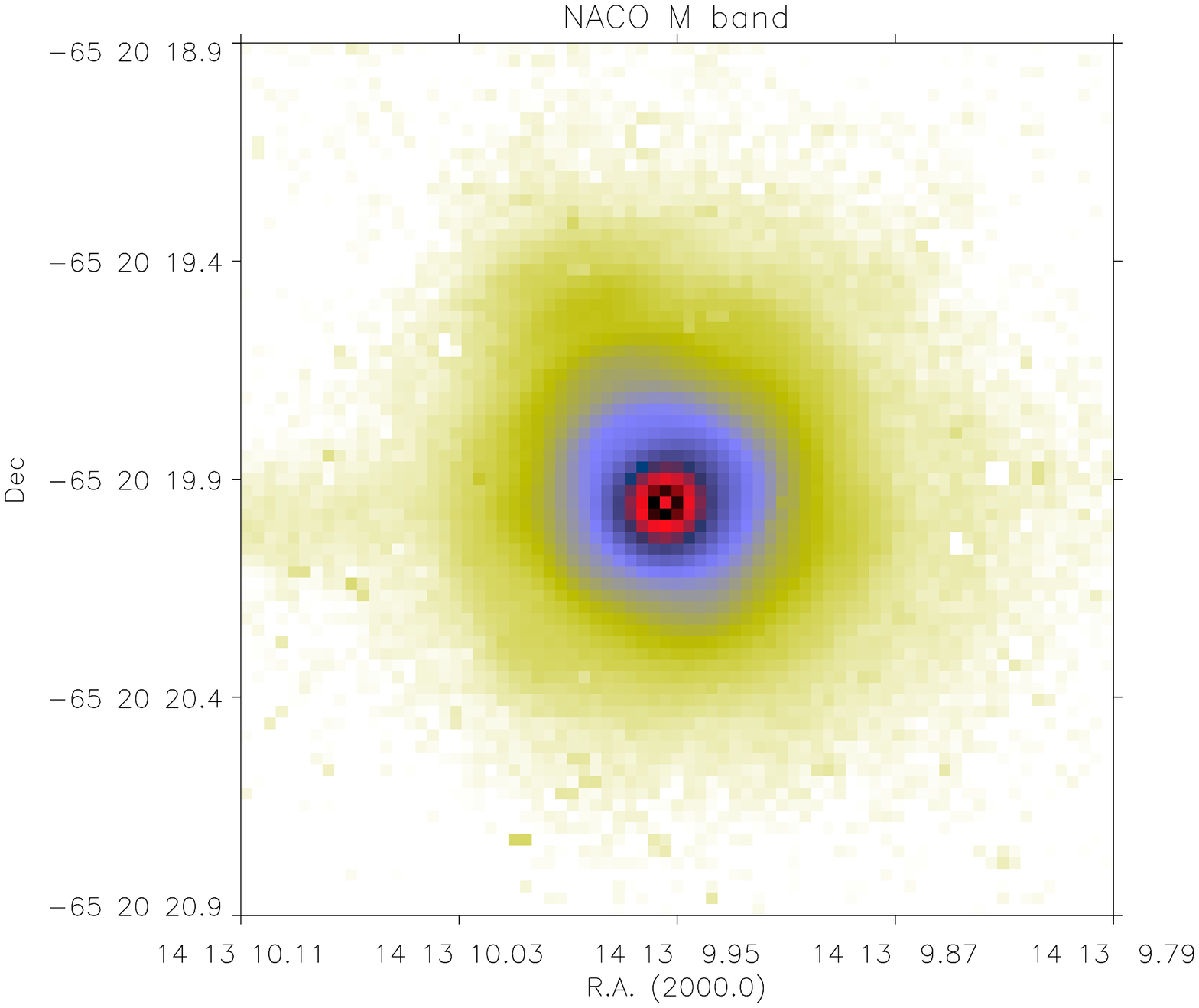,width=65mm}\hspace{15mm}\psfig{file=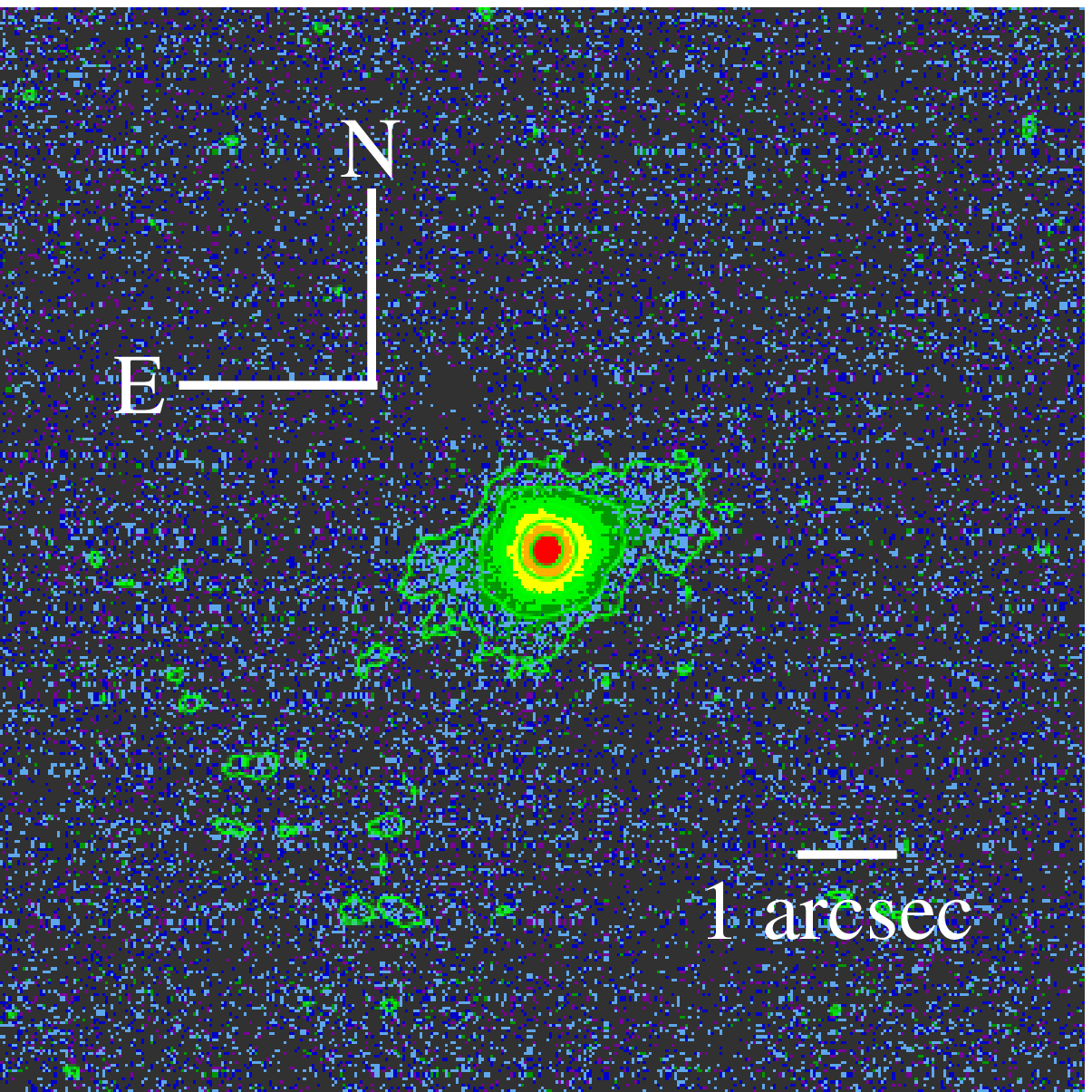,width=50mm}\hspace{5mm}}

\caption{HST and NACO images of the central
  $20\arcsec\times20\arcsec$ region of Circinus
  ($2\farcs5\times2\farcs5$ in M\p-band): 
Top left F814W (8140\AA) WFPC2-PC chip;
Top right J-band;
Center left 2.42\,\micron;
Center right L\p-band;
Bottom left M\p-band;
Bottom right pure coronal [Si VII]~2.48\,$\mu$m line emission.
}
\label{all}
\end{figure}
\clearpage


\begin{figure}
\epsscale{1.0}
\plotone{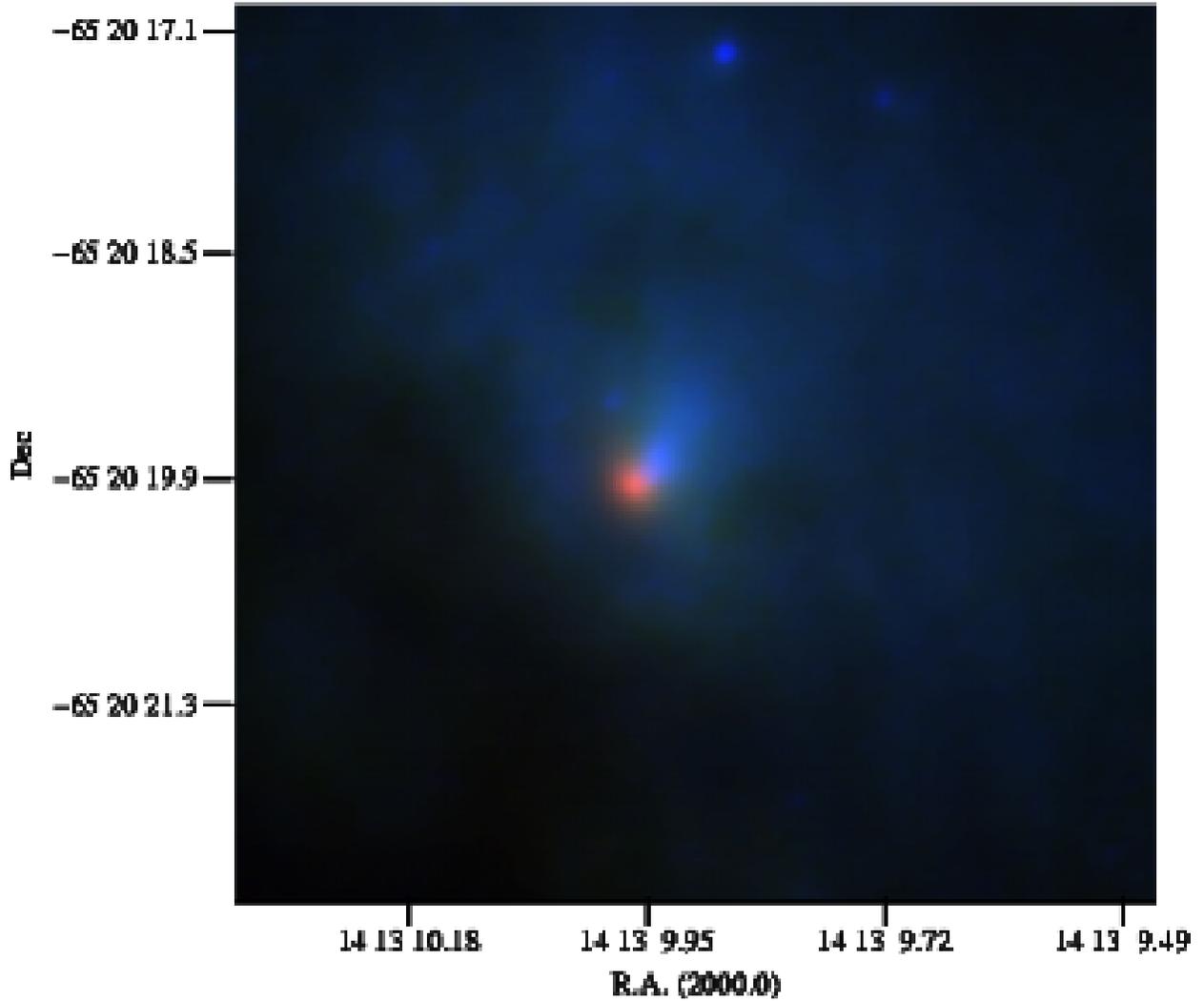}
\caption{ True color image combining HST WFPC2-PC F814W (blue) 
and NACO J- (green)
and Ks-bands (red). 
The figure shows the 10 pc collimated beam (to which only the
optical and J-band contribute) pointing away from the
bright central source (comprising only the K-band emission) which
we argue is the nucleus.
}
\label{col}
\end{figure}

\begin{figure}
\plotone{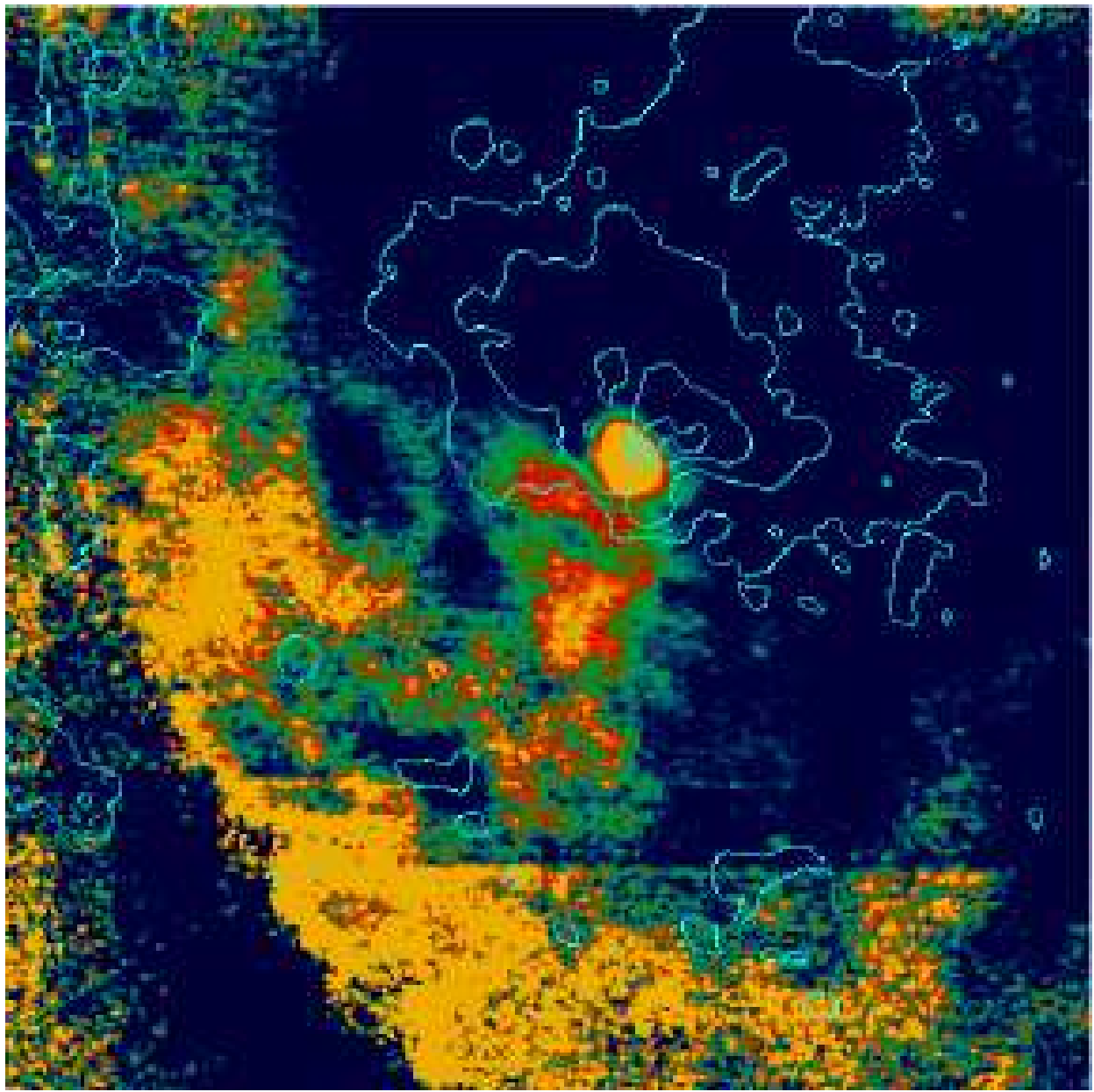}
\caption{NACO [J-2.42\,$\mu$m] color image.
The field of view is $20\arcsec\times20\arcsec$, centered on the nucleus.
North is up and East is left. 
The superimposed contours trace the HST H$\alpha$ emission.
This shows Circinus' one-sided ionization cone to the north-west as
well as some of the brightest star forming regions from Circinus' circumnuclear ring, which coincide
with regions of lower extinction, in in
the eastern and southern parts of the galaxy.
}
\label{jk}
\end{figure}

\begin{figure}
\epsscale{1.0}
\plotone{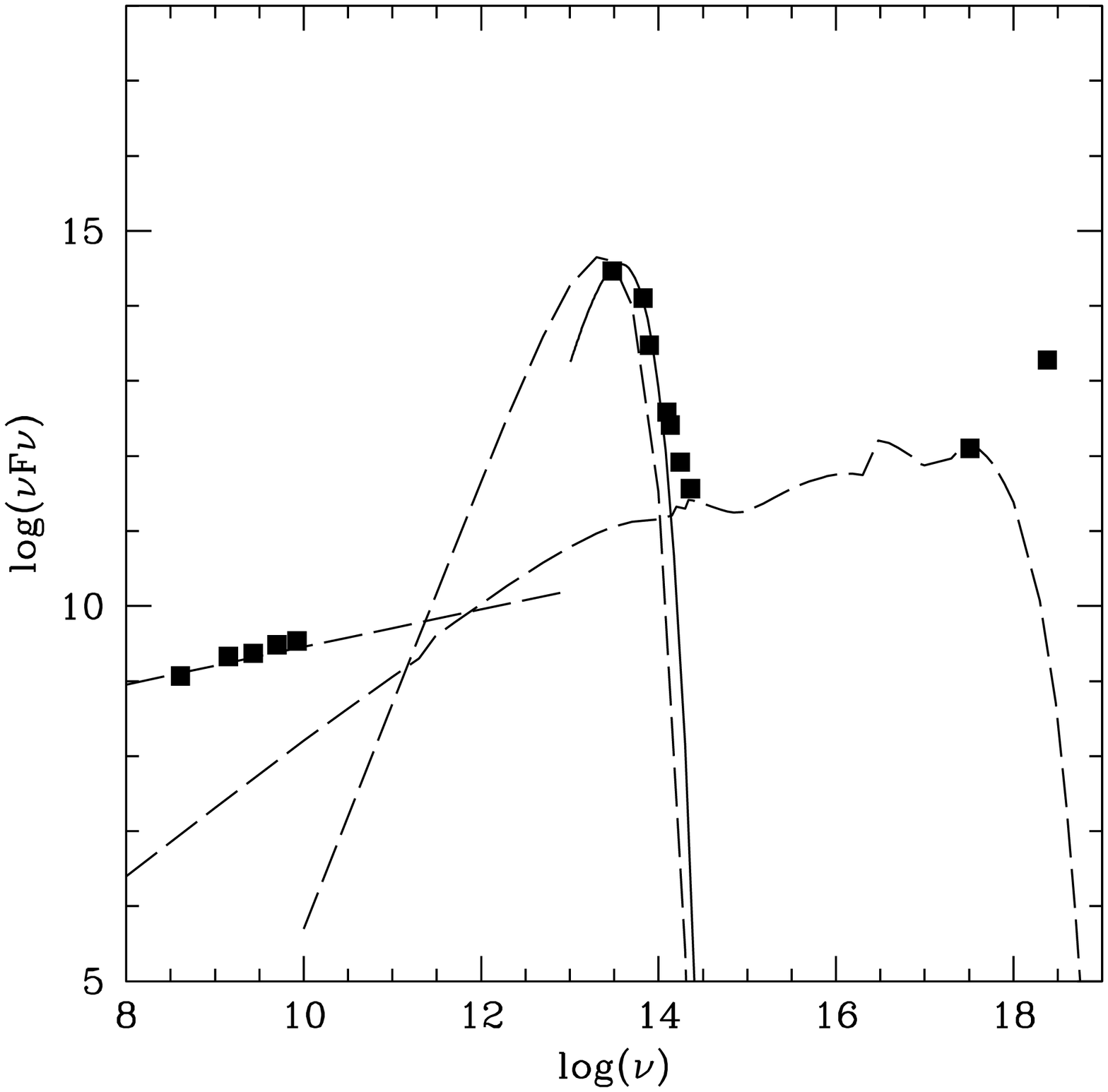}
\caption{SED of Circinus' nucleus.
The new infrared data (in the range $\log{\nu}=13.5$--14.5) are from 
Table 1. Note that the J-band flux is an upper limit to the infrared core
source, since no counterpart to this source is seen shortwards
of 1.3\,$\mu$m.
The continuous line is a modified black-body function with
T=300\,K fit to the infrared SED; 
The dashed line  is a radiation transfer model of the complete
SED comprising three components: dust reradiation 
accounting for  the infrared,
bremsstrahlung from cooling gas dominating the UV to X-ray regime, and
synchrotron emission dominating the radio flux.}
\label{planck}
\end{figure}




\newpage
 


\begin{thebibliography}{}

\bibitem[Rodriguez-Ardila]{rod04}
Rodriguez-Ardila, Prieto \& Viegas 2004, in ``The Interplay among Black Holes,Stars and ISM in Galactic Nuclei, Proceedings IAU Symposium No. 222 in press.


\bibitem[Brinkmann]{bri94}
Brinkmann, W., Sibert, J \&  Boller, Th. 1994, A\&A 281, 355

\bibitem[Contini, Viegas \& Prieto 2003]{con03}
Contini, M, Viegas, S.M. \& Prieto, M.A.  2004, MNRAS 348, 1065

\bibitem[]{u1}
Contini, M, Viegas, S.M. \& Prieto, M.A.  1998, ApJ 505, 621 

\bibitem[]{u2}
Elmouttie, M.,  Haynes, R., Jones, K., Ehle, M., Beck, R. \& Wielebinski, R.
1995, MNRAS 275, L53

\bibitem[]{u3}
Freeman, K.C.,  Karlsson, B., Linga, G., Burrell, J. F., van Woerden, H. 
\& Goss W. M. 1977, A\&A 55, 445


\bibitem[]{u4}
Giovanardi, C., Hunt, L. K. 1996, AJ, 111, 1086 

\bibitem[Greenhill et al. 2003]{gre03}
Greenhill et al. 2003, ApJ 582, L11

\bibitem[]{u5}
Heijligers, H., 2003, private communication


\bibitem[]{u6}
Quillen, A. et al. 2001, ApJ 547, 129


\bibitem[]{u7}
Maiolino, R., ALonso-Herrero, A., Andres, S., Quillen, A., Rieke, M.,
Rieke, G. H., Tacconi-Garman, L. 2000,  ApJ 531, 219

\bibitem[Marconi et al. 1994]{mar94}
Marconi, A.  et al. 1994, Messenger 78, 20

\bibitem[Matt et al. 1999]{mat99}
Matt, G., et al. 1999, A\&A 341, L39

\bibitem[]{u8}
Smith, D.A., Wilson, A.S. 2001, ApJ, 557, 180 

\bibitem[Wilson et al. 2000]{wil00}
Wilson, A. S., Shopbell, P. L., Simpson, C., Storchi-Bergmann, T.,
Barbosa, F. K. B., \& Ward, M. J. 2000, \apj, 120, 1325

\end{thebibliography}
\end{document}